\documentstyle[12pt]{article}
\title{Quantum properties of classical Fisher information}
\author{Michael J. W. Hall\\ 
Theoretical Physics, IAS\\Australian National University\\
Canberra ACT 0200, Australia}
\date{}
\begin{document}
\maketitle

\begin{abstract}
The Fisher information of a quantum observable is shown to be proportional to 
both (i) the difference of a quantum and a classical variance, thus providing a measure of 
nonclassicality; and (ii) the rate of entropy increase under Gaussian diffusion, thus providing 
a measure of robustness.  The joint nonclassicality of position and momentum observables is 
shown to be complementary to their joint robustness in an exact sense.

\end{abstract}

\renewcommand{\thesection}{\Roman{section}}
\renewcommand{\thesubsection}{\Alph{subsection}}

\newpage

\section{INTRODUCTION}

Fisher information was originally introduced by Fisher in 1925 \cite{fish}, as a measure of  
``intrinsic accuracy'' in statistical estimation theory.  It provides in
particular a bound on the
degree to which members of a family of probability distributions can be distinguished 
\cite{cox}.  Quantum generalizations of Fisher information may be given, 
providing corresponding bounds on the degree to which members of a family of quantum 
states can be distinguished by measurement \cite{helhol}.  However, in both the classical 
and quantum contexts, the bounds are typically not achievable.  Hence the primary 
application of the Fisher information has been in providing unsharp statistical inequalities.

In this paper two curious connections between {\it classical} Fisher information and {\it 
quantum} systems will be pointed out, which involve exact {\it equalities}.  First, it is shown 
that the classical Fisher information of a quantum observable is proportional to the difference 
between the quantum variance and the classical variance of the conjugate observable.  Thus 
it is a direct measure of the {\it nonclassicality} of the conjugate observable.  Second, it is 
shown that the classical Fisher information is proportional to the rate of entropy increase of 
the observable when the quantum system is subjected to Gaussian diffusion, ie, Brownian 
motion.  Hence it is also a measure of the {\it robustness} of the observable with 
respect to noise.  The results further lead to natural measures of joint nonclassicality and 
joint robustness for quantum states, which are inversely related to each other.

Fisher information is defined in the following Section, and its relation to statistical measures 
of uncertainty briefly reviewed.  In Sec. III the role of Fisher information as a measure of 
nonclassicality is developed and explored, based on a natural
decomposition of each quantum observable into a ``classical'' and a
``nonclassical'' component.  The ``joint nonclassicality'' 
of a quantum system 
is defined, and it is conjectured that it has a nontrivial lower bound for pure quantum states, 
i.e., such states are inherently nonclassical.   In Sec. IV the 
connection between Fisher information and quantum diffusion is demonstrated, essentially 
generalizing de Bruijn's identity for classical systems \cite{stam, cov1,cov2}.  It follows that 
the robustness of a quantum system with respect to noise is inversely proportional to its 
degree of nonclassicality; i.e.,  the more robust the state is with
respect to noise, the more classical it is. Generalizations 
to higher dimensions are briefly discussed in Sec. V,  
and conclusions given in Sec. VI.

\section{FISHER INFORMATION AND FISHER LENGTH}

The classical Fisher information associated with translations of a one-dimensional observable 
$X$ with corresponding probability density $p(x)$ is given by \cite{fish,cox}
\begin{equation} \label{fish}
F_X = \int dx p(x) [d \ln p(x)/dx]^2  > 0 .
\end{equation}
The primary application of this quantity in classical estimation theory is the 
lower bound
\begin{equation} \label{cr1}
Var X \geq F_X^{-1} 
\end{equation}
 for the variance of $X$, known as the Cramer-Rao inequality
\cite{cox}.

One may also define a corresponding Fisher {\it length} for $X$, by
\begin{equation} \label{fishlen}
\delta X = F_X^{-1/2}  .
\end{equation}
From Eq. (\ref{fish}) $\delta X$ is seen to quantify the length scale over which $p(x)$ (or 
more precisely $\ln p(x)$) varies appreciably.  The Cramer-Rao inequality Eq. (\ref{cr1}) 
may then be rewritten as the simple length inequality
\begin{equation} \label{cr2}
\Delta X \geq \delta X
\end{equation}
for the root mean square deviation $\Delta X$ of $X$.

It is worth noting that Eq. (\ref{cr2}) 
can be derived via the properties of a length measure of fundamental 
geometric significance, the {\it ensemble length} of $X$, 
given by the exponential of the entropy of $p(x)$:
\begin{equation} \label{lx}
L_X = \exp [-\int dx p(x) \ln p(x)]  .
\end{equation}
$L_X$ 
is the unique measure of uncertainty that satisfies several basic geometric 
properties expected of a ``length'' \cite{hall}, and one has 
\begin{equation} \label{ineq}
\sqrt{2\pi e} \Delta X \geq L_X \geq \sqrt{2\pi e} \delta X . 
\end{equation}
The first inequality in Eq. (\ref{ineq}) corresponds to 
the well known property that entropy is 
maximised for a fixed value of $\Delta X$ by a Gaussian distribution.  The second inequality 
may be derived from either an identity of de Bruijn \cite{cov1,cov2} (see also Sec. IV), or 
from a logarithmic Sobelov inequality \cite{carl}, and is also saturated by  Gaussian 
distributions.  The Cramer-Rao inequality Eq. (\ref{cr2}) immediately
follows from Eq. (\ref{ineq}).

\section{MEASURE OF NONCLASSICALITY}

\subsection{Position}
Consider a quantum system described by wavefunction $\psi (x)$.  The position probability 
density is then $p(x)=|\psi (x)|^2$, and hence from Eq. (\ref{fish}) the corresponding Fisher 
information is
\begin{eqnarray}
F_X & = & \int dx |\psi(x)|^2[\psi '(x)/\psi (x) + \psi^*{'}(x)/\psi^*
(x) ]^2 \nonumber \\
& = & 4\int dx \psi^*{'} (x)\psi'(x) + \int dx 
|\psi(x)|^2[\psi '(x)/\psi (x) - \psi^*{'}(x)/\psi^*
(x) ]^2
\nonumber \\ \label{diff}
& = & (4/\hbar )^2 \left[ \langle P^2\rangle_\psi - \langle P_{cl}^2\rangle_\psi \right]  ,
\end{eqnarray}
where $P$ denotes the momentum observable conjugate to $X$, and $P_{cl}$ is a {\it 
classical} momentum observable
corresponding to the state $\psi$, given by the function
\begin{equation} \label{pcl}
P_{cl} (x) = (\hbar/2i)[\psi '(x)/\psi (x) - \psi^*{'}(x)/\psi^*(x)]  .
\end{equation}

The identification of the observable $P_{cl}$ with a classical momentum is strongly 
supported on two grounds.  First, the probability density $|\psi (x)|^2$ is well known to 
satisfy the classical continuity equation \cite{merz}
\begin{equation} \label{cont}
\partial |\psi (x)|^2/\partial t + (\partial /\partial x) \left[
|\psi (x)|^2 m^{-1} P_{cl}(x)\right] =0  ,
\end{equation}
as a direct consequence of the Schr\"{o}dinger equation.  Thus $ m^{-1} P_{cl}(x)$ is the 
local velocity of probability flow in position space, implying $P_{cl}(x)$ may be interpreted 
as  a classical momentum of a particle at position $x$, where the probability of finding the 
particle at $x$ is $|\psi (x)|^2$.  Second, one has the identity 
\begin{equation} \label{expec}
\langle P \rangle_\psi = \langle
P_{cl} \rangle_\psi 
\end{equation}
following
from Eq. (\ref{pcl}) (using integration by parts). 
Hence the expectation values of the 
observables $P$ and $P_{cl}$ are equal for all 
wavefunctions. 

Now, given the quantum and classical momentum observables $P$ and $P_{cl}$, it is 
natural define the {\it nonclassical} momentum of the system
by $P_{nc}=P-P_{cl}$.  Thus the momentum $P$ separates into a classical and
a nonclassical contribution.  From Eq. (\ref{pcl}) one has
\begin{eqnarray*}
\langle PP_{cl} +P_{cl}P\rangle_\psi & = & \int dx (P\psi)^*(P_{cl}\psi)
+ \int dx (P_{cl}\psi)^*(P\psi) \\
& = & (\hbar/i) \int dx[\psi
'(x)\psi^*(x)-\psi^*{'}(x)\psi(x)]P_{cl}(x)\\
& = & 2 \langle P_{cl}^2\rangle_\psi  ,
\end{eqnarray*}
and so from Eq. (\ref{expec}) (with $p=0$) 
\begin{equation} \label{var}
Var_\psi P = Var_\psi P_{cl} + Var_\psi P_{nc} .
\end{equation}
Hence the classical and nonclassical contributions are uncorrelated in
variance.

The main result of
this section is a simple relationship between nonclassicality and
Fisher information.  In particular, from
Eqs. (\ref{diff}), (\ref{expec}) and (\ref{var}) one has 
\begin{equation} \label{vardiff}
F_X = (4/\hbar^2) (\Delta P_{nc})^2 . 
\end{equation}
The position Fisher information is therefore proportional to the nonclassical
variance of the conjugate momentum.

A {\it direct} measure of the nonclassicality of the momentum,
representing the size of nonclassical momentum fluctuations, is
given
by the root mean square deviation $\Delta P_{nc}$.
From Eq. (\ref{fishlen}) one may equivalently write Eq.
(\ref{vardiff}) as
\begin{equation} \label{preheis}
\delta X \Delta P_{nc} = \hbar /2  .
\end{equation}
Thus {\it the Fisher length of position is inversely 
proportional to the nonclassicality of momentum}.
Eq. (\ref{preheis}) is rather
similar
 in form to the
Heisenberg uncertainty relation, and indeed the latter may be
immediately derived from it.
In particular, one has
\begin{equation} \label{heis}
\Delta X \Delta P \geq \delta X \Delta P \geq \delta X \Delta P_{nc} =\hbar/2  ,
\end{equation}
where the first inequality follows from Eq. (\ref{cr2}), and the second
from Eq. (\ref{var}).  The inequality $\delta X\Delta P\geq\hbar/2$
implicit in Eq. (\ref{heis}) was first proved by Stam
\cite{stam,cov2}, based on a Schwarz inequality.

Note that the Fisher length is always 
finite from Eqs. (\ref{fish}) and (\ref{fishlen}), 
and hence the momentum 
nonclassicality is never zero.  Further, from Eq. (\ref{var}), the 
momentum nonclassicality is maximum, for a fixed value 
of $\Delta P$, when the variance of 
$P_{cl}$ vanishes, i.e., when $P_{cl}$ is a constant.  From Eq. (\ref{pcl}) this occurs when 
the phase of $\psi (x)$ is linear in $x$.  Thus
\begin{equation} \label{stam}
\Delta P_{nc} = \Delta P \mbox{  iff  } \arg \psi (x) = 
\alpha + p_0x  ,
\end{equation}
for constants $\alpha$ and $p_0$.  

\subsection{Momentum}

One may, in direct analogy with Eqs. (\ref{var}),
(\ref{vardiff}) and (\ref{preheis}), obtain the conjugate 
equalities
\begin{equation} \label{fp}
F_P = (4/\hbar^2) (\Delta X_{nc})^2 = (4/\hbar^2) [Var_\psi X - Var_\psi X_{cl}] ,
\end{equation}
\begin{equation} \label{momp}
\Delta X_{nc} \delta P =\hbar/2  ,
\end{equation}
for the Fisher information $F_P$ and the Fisher length $\delta P$ of the 
momentum observable $P$ conjugate to $X$.  Here $X_{nc}=X-X_{cl}$,
and 
\begin{equation} \label{xcl}
X_{cl}(p) = (i\hbar /2) [\phi 'p)/\phi(p) - \phi^*{'}p)/\phi^*(p)]
\end{equation}
is a classical position observable corresponding to state $\psi$, where $\phi(p)$ denotes the 
momentum wavefunction of the system.
Thus $F_P$ and $\delta P$ are related to the nonclassicality of the
position.

The identification of $X_{cl}$ as a classical position observable has a similar justification to 
the analogous interpretation of $P_{cl}$.  In particular, conservation
of momentum probability $|\phi(p)|^2$ implies a 
continuity equation of the form 
\begin{equation} \label{contp}
\partial |\phi(p)|^2/\partial t + (\partial /\partial p) \left[ 
|\phi(p)|^2 F(p)\right] = 0  , 
\end{equation}
where $F(p)$ is the momentum flow, i.e.,  {\it force },
associated with momentum $p$.  If the system is subject to a potential
energy $V(x)$, then multiplying the Schr\"{o}dinger equation 
in the momentum representation by
$\phi^*(p)$, taking the imaginary part, and expanding $V(x)$ in a Taylor
series, one finds
\begin{equation} \label{fcl}
F(p)  = - (\partial/ \partial x) V(X_{cl}(p)) + O(\hbar^3) .
\end{equation}
Thus the observable $X_{cl}$ corresponds to the classical 
force $-V'(x)$ associated with the system, 
at least to second order in $\hbar$.  
One has also an equality analogous to Eq. (\ref{expec}), i.e., 
\begin{equation}
\langle X\rangle_\psi = \langle 
X_{cl} \rangle_\psi  .
\end{equation}

\subsection{Joint nonclassicality}

A natural (dimensionless) measure of {\it joint} nonclassicality for a quantum state $\psi$ 
may now be defined, as
\begin{equation} \label{joint}
J_{nc} = \Delta X_{nc}\Delta P_{nc}/(\hbar/2)  .
\end{equation}
From Eqs. (\ref{preheis}) and (\ref{momp}) one then has 
\begin{equation} \label{joint2}
J_{nc} = (\hbar/2) (\delta X \delta P)^{-1} ,
\end{equation}
i.e., {\it the joint nonclassicality is inversely 
proportional to the product of the position and 
momentum Fisher lengths}.  Recalling that equality holds throughout 
Eq. (\ref{ineq}) for Gaussian 
distributions, it follows that $J_{nc}=1$ for minimum uncertainty states.

It is of interest to ask whether there is some maximum {\it upper} bound for joint 
nonclassicality set by quantum theory, corresponding to a lower bound for the product 
$\delta X\delta P$.  The answer is in the negative; in particular, there is no direct analogue of 
the Heisenberg uncertainty relation Eq. (\ref{heis}) for Fisher lengths.  As an example, 
consider the the $n$th energy eigenstate of a one-dimensional harmonic oscillator.  For this 
case the momentum and position wavefunctions are 
both real up to a constant phase factor. 
Hence from Eq. (\ref{stam}) and its analogue for the momentum wavefunction, 
\begin{equation} \label{noheis}
\delta X \delta P = (\hbar^2/4) (\Delta X \Delta P)^{-1} = \hbar /(4n+2) ,
\end{equation}
which becomes arbitrarily small as $n\rightarrow\infty$.  
Thus the joint nonclassicality 
becomes arbitrarily large with increasing $n$.  Note 
that the first equality in Eq. (\ref{noheis}) holds whenever the 
position wavefunction is (up to a linear phase factor and a translation) 
a symmetric or 
antisymmetric real function.

The definition of joint nonclassicality may be extended to mixed states, represented by 
density operators, via Eqs. (\ref{fish}), (\ref{fishlen}) and (\ref{joint2}).   For such states 
$J_{nc}$ can be arbitrarily small (e.g., consider thermal states of the harmonic oscillator, 
which have Gaussian position and momentum distributions, in the high temperature limit).  
This is reasonable, as one expects certain mixed states, such as thermal states, to be 
equivalent to classical states in appropriate limits.  However, it would be of interest to 
determine whether there is a {\it non-zero} minimum value for the joint nonclassicality of 
{\it pure} states.  This would correspond to the idea that there is 
necessarily something inherently nonclassical about a pure quantum
state.  The general uncertainty relation $\Delta A\Delta B\geq|\langle
[A,B]\rangle_\psi |/2$ \cite{merz} implies via Eq. (\ref{joint}) that
\begin{equation} \label{bound}
J_{nc}\geq |1 + (i/\hbar) \langle [P_{cl}, X_{cl}]\rangle_\psi |,
\end{equation}
suggesting the conjecture $J_{nc}\geq 1$ for pure states. 

\subsection{Kinetic energies and quantum potentials}

From Eq. (\ref{diff}) it is seen that the position 
Fisher information $F_X$ is proportional to the 
difference of a quantum and a classical kinetic energy.  
Thus the average energy of a 
quantum particle of mass $m$ is increased relative to the corresponding 
average classical energy by 
the additional amount
\begin{equation} \label{energy}
E_F = \hbar^2 F_X  /(8m) .
\end{equation}
Now, it is known from the de Broglie-Bohm approach to quantum mechanics that 
there is an {\it exact} correspondence between a quantum particle and
an ensemble of classical particles, where the latter has probability 
density $p(x)=|\psi (x)|^2$, momentum $P_{cl}(x)$ associated with
position $x$, and is subjected 
to a quantum potential $Q(x)$ in addition to the classical 
potential $V(x)$, where \cite{bohm} 
\begin{equation} \label{qx}
Q(x) = \hbar^2/(8m) [p'(x)^2/p(x)^2 - 2p''(x)/p(x)]  . 
\end{equation}
The average energy increase due to $Q(x)$ is therefore $\langle
Q(x)\rangle_\psi$, and hence from Eq. (\ref{energy}) one has
\begin{equation} \label{en}
\langle Q(x)\rangle_\psi = \hbar^2 F_X  /(8m) . 
\end{equation}
Thus $F_X$ is proportional to the average value of the quantum potential,
providing another link between Fisher information and nonclassicality.

Eq. (\ref{en}) was recently derived by Reginatto \cite{reg1} 
based on an even stronger connection 
between the quantum potential and Fisher information.  
In particular, consider 
the variation of $F_X$ in Eq. (\ref{fish}) with respect to the probability density $p(x)$.  
One then finds, using integration by parts, the remarkable relation 
\begin{equation} \label{variation}
\delta F_X = (8m/\hbar^2) \int dx Q(x) \delta p .
\end{equation}
This result is the basis of a new approach to quantum mechanics, where a Fisher information 
term and a classical hydrodynamical action term are added, representing 
``epistemological'' and ``ontological'' contributions respectively to the
total action
\cite{reg1,reg2}.   This approach is to be distinguished from 
that of Frieden \cite{fried}, where essentially a {\it generalized} 
Fisher information is defined for wavefunctions, 
proportional to the quantum kinetic energy $\langle 
P^2/(2m)\rangle_\psi$.

\section{MEASURE OF ROBUSTNESS}

\subsection{De Bruijn's identity}

Consider the entropy increase of an observable $X$ subjected to Gaussian
diffusion, i.e., Brownian motion.
The probability density $p_t(x)$ satisfies the diffusion equation
\begin{equation} \label{brown}
\dot{p}_t = \gamma p_t''
\end{equation}
for some diffusion rate constant $\gamma$, with solution
\begin{equation}  \label{conv}
p_t(x) = (\pi\gamma t)^{-1/2}\int dy p_0(x-y)\exp [-y^2/(\gamma t)]  ,
\end{equation}
and hence the initial density is convolved with a Gaussian of variance
$\gamma t
/2$.  The rate of entropy
increase at time $t$ is therefore given by
\begin{equation} \label{rate}
\dot{S}_X(t)  =  -\int dx \left[ 1 + \ln p_t(x)\right] \dot{p}_t(x)
= \gamma F_X(t)  ,
\end{equation}
where $F_X(t)$ is the Fisher information at time $t$ and the second 
equality follows
 from Eqs. (\ref{fish}) and  
(\ref{brown}), using integration by parts.

This link between Fisher information and entropy increase is known as
{\it de Bruijn's identity}
\cite{stam,cov1,cov2}.  Since an observable which is robust to noise
will have a
 small rate of entropy
increase, and vice versa, it follows that $F_X=F_X(0)$ is inversely
related to the
{\it robustness} of $X$
with respect to the onset of Gaussian noise.

The application of de Bruijn's identity to quantum systems is
straightforward.  
Moreover, even though the
position and momentum observables are complementary, and hence cannot be
specified simultaneously, it
turns out that these observables behave independently when the system is
subjected
to simultaneous position
and momentum diffusion.  Hence the quantum analogues of Eq. (\ref{rate})
for position and momentum can
be derived from a {\it single} quantum diffusion process.

In particular, the diffusion equation for a classical phase
space ensemble $\rho (x,p)$ is
\begin{eqnarray*}
\dot{\rho} = \gamma (\partial^2/\partial x^2)\rho + \sigma
(\partial^2/\partial
p^2)\rho
& = & \gamma \{p,\{p,\rho\}\} + \sigma \{x,\{x,\rho\}\}  ,
\end{eqnarray*}
where $\gamma$ and $\sigma$ are rate constants and $\{ ,\}$ is the
Poisson bracket.  Under the Dirac
correspondence $\{ ,\}\rightarrow (i\hbar )^{-1}[,]$ one thus obtains
the quantum
diffusion equation
\begin{equation} \label{master}
\dot{\rho} = -(\gamma/\hbar^2)[P, [P, \rho]]
- (\sigma/\hbar^2) [X,[X,\rho]]  ,
\end{equation}
where $\rho$ is the density operator describing the system
\cite{footnote}. 

The evolution of the position probability density $p_t(x)=\langle x|\rho
|x\rangle$ is therefore given by
\begin{equation}
\dot{p}_t(x) = -(\gamma/\hbar^2) \langle x|[P,[P,\rho]]|x\rangle -
(\sigma/\hbar
^2) \langle
x|[X,[X,\rho]]|x\rangle  .
\end{equation}
The second term on the right vanishes since $X|x\rangle = x|x\rangle$ by
definition, while the first term
reduces to $\gamma p_t''(x)$ using the relation $\langle x|[P,A]|x\rangle
= (\hbar
/i)d\langle x|A|x\rangle /dx$
(derived by expanding $|x\rangle$ in momentum eigenstates). Thus 
$p_t(x)$ satisfies the diffusion
equation Eq. (\ref{brown}).  A similar result obtains for the evolution
of the momentum density, and hence
from de Bruijn's identity Eq. (\ref{rate}) one has
\begin{equation} \label{quant}
F_X = \gamma \dot{S}_X(0) , \mbox{     } F_P = \sigma \dot{S}_P(0)  .
\end{equation}
Thus, the position and momentum Fisher informations of a quantum system
are inversely related
to the robustness of the corresponding observables, with respect to the
onset of
 quantum phase space
diffusion as per Eq. (\ref{master}). 

Noting Eqs. (\ref{vardiff}), (\ref{fp}) and (\ref{quant}), 
the robustness of the position
is high (small $F_X$) when the nonclassicality of the momentum is low,
and vice
versa.  Thus {\it the more
classical an observable is, the more robust the conjugate observable is
with respect to noise}.

\subsection{Joint robustness}

From Eqs. (\ref{fishlen}) and (\ref{quant}) a natural (dimensionless)
measure of {\it joint} robustness for a
quantum system is given by
\begin{equation} \label{jr}
J_r = \delta X \delta P /(\hbar/2)  .
\end{equation}
In particular, $J_r$ is relatively {\it large} when the position and
momentum entropies increase {\it
slowly} under the onset of phase space diffusion, and vice versa.  Note
from Eq.
(\ref{noheis}) that the joint
robustness can be arbitrarily small.  Conversely, thermal states of the
harmonic
 oscillator have arbitrarily
large robustness in the high temperature limit.

Comparison of Eqs. (\ref{joint2}) and (\ref{jr}) shows that
\begin{equation} \label{inverse}
J_{nc} J_r = 1  ,
\end{equation}
i.e., {\it the nonclassicality and robustness of a quantum state are
inversely proportional}.  This is in
accord with other results in the literature suggesting that classical
behaviour
is associated with robustness to
noise \cite{decoh}.

\section{HIGHER DIMENSIONS}

In more than one dimension the Fisher information generalizes to a
matrix.  However, the results of the
previous sections can and do generalize in different ways, to relations
involving either the matrix, its trace,
or its determinant (which are all equivalent in one dimension).  It is
therefore
 useful to indicate explicitly the
higher-dimensional analogs of various results.

First, for an $n$-dimensional observable ${\bf X}$ with probability
density $p({
\bf x})$, the analog of the
Fisher information in Eq. (\ref{fish}) is the (positive definite) Fisher
matrix
\cite{cox}
\begin{equation} \label{fishmat}
F_{\bf X} = \int d^n{\bf x} p({\bf x})
[\nabla \ln p({\bf x})] [\nabla\ln p({\bf
x})]^T  ,
\end{equation}
where $\nabla$ is the gradient operator and $T$ denotes the vector
transpose.  The
Cramer-Rao inequality
Eq. (\ref{cr1}) then generalizes to \cite{cox}
\begin{equation} \label{crmat}
Cov({\bf X}) = \langle {\bf  X X}^T\rangle - \langle {\bf
X}\rangle\langle{\bf
X}^T\rangle \geq F_{\bf
X}^{-1}
\end{equation}
for the covariance matrix of ${\bf X}$, and the inequality chain in Eq.
(\ref{ineq}) becomes
\begin{equation} \label{ineqmat}
(2\pi e)^{n/2} \Delta V \geq V_{X} \geq (2\pi e)^{n/2} \delta V
\end{equation}
for the root mean square volume $\Delta V=[\det Cov({\bf X})]^{1/2}$ and
Fisher
volume $\delta V=[\det
F_{\bf X}]^{-1/2}$, where $V_X$ denotes the ensemble volume, 
given by the exponential of the ensemble entropy \cite{hall}.
The first inequality corresponds to the variational property that
entropy
 is maximized for  a given
covariance by a Gaussian distribution, and the second inequality is
given by Dembo et al. (Sec. IV.C of
\cite{cov2}).

For a quantum system described by wavefunction $\psi({\bf x})$ one
finds, in analogy to Eqs. (\ref{var})
and (\ref{vardiff}),
\begin{equation} \label{covmat}
F_{\bf X} = (4/\hbar^2) \left[ Cov_\psi({\bf P}) - Cov_\psi({\bf
P}_{cl})\right] = (4/\hbar^2) Cov_\psi({\bf
P}_{nc}) ,
\end{equation}
where ${\bf P}_{cl}$ is a classical momentum vector defined by replacing
$\psi '(x)$
 by $\nabla\psi({\bf x})$
in Eq. (\ref{pcl}), and ${\bf P} = {\bf P}_{cl} + {\bf P}_{nc}$.   Hence
{\it the
 position Fisher matrix is
proportional to the covariance matrix of the nonclassical momentum}.  A
conjugate relation holds for
$F_{\bf P}$.

There are two natural scalar measures of joint nonclassicality which
reduce to
the measure in Eq.
(\ref{joint}) for $n=1$.  The first is
\begin{equation} \label{jointmat1}
J_{nc}^{(1)}  = (\hbar/2)^{-n}[\det Cov_\psi({\bf X}_{nc})\det
Cov_\psi({\bf P}_{nc})]^{1/2} =
(\hbar/2)^{n} [\det F_{\bf X} \det F_{\bf P}]^{1/2} ,
\end{equation}
which may be interpreted as a (dimensionless) nonclassical phase space
volume.
The second is
\begin{equation} \label{jointmat2}
J_{nc}^{(2)} = (\hbar/2)[tr F_{\bf X} tr F_{\bf  P}]^{1/2}  ,
\end{equation}
which turns out to be related to joint robustness in the general case
(see below
).

The analog of Eq. (\ref{diff}) is
\begin{equation} \label{diffmat}
tr F_{\bf X} = (4/\hbar^2) \left[\langle {\bf P.P}\rangle_\psi -
\langle {\bf P}_{cl}{\bf .P}_{cl}\rangle_\psi \right]  ,
\end{equation}
and thus the trace of the Fisher matrix is proportional to the
difference of a quantum and a classical kinetic
energy (and hence to the average of the quantum potential energy as per
Sec. III
.D).  Also, from Eq.
(\ref{covmat}) one has the Stam inequality $Cov({\bf P})\geq
(\hbar^2/4)F_{\bf X
}$ \cite{stam,cov2},
which multiplied by Eq. (\ref{crmat}) yields the Heisenberg uncertainty
relation
\begin{equation} \label{heismat}
Cov({\bf X}) Cov({\bf P}) \geq (\hbar^2/4) I_n  ,
\end{equation}
where $I_n$ denotes the $n\times n$ identity matrix.

Finally, the de Bruijn identities in Eq. (\ref{quant}) generalize  to
give a somewhat less direct connection
between the Fisher matrix and entropy increase in higher dimensions.  In
particular consider the $n$-dimensional
analog of the diffusion equation Eq. (\ref{brown}),
\begin{equation} \label{brownmat}
\dot{p}_t = \left( \nabla^T \Gamma\nabla\right) p_t  ,
\end{equation}
where $\Gamma$ denotes a real symmetric positive diffusion matrix
with constant coefficients.  Under
the coordinate transformation ${\bf y}=\Gamma^{-1/2}{\bf x}$ this
reduces to the
 canonical form
$\dot{p}_t=\nabla^2p_t$, and in exact analogy to  the derivation of Eq.
(\ref{rate}) one finds that the {\it
trace} of the Fisher matrix for ${\bf Y}$ is equal to the rate of
entropy increase of ${\bf Y}$.  But it follows
directly from the coordinate transformation that $F_{\bf
Y}=\Gamma^{1/2}F_{\bf
X}\Gamma^{1/2}$ and
$S_{\bf Y}=S_{\bf X}-(1/2)\ln\det\Gamma$, and hence one has the
generalization
\begin{equation} \label{ratemat}
\dot{S}_{\bf X}(t) = tr [\Gamma F_{\bf X}(t)]
\end{equation}
of Eq. (\ref{rate}).

For the case of isotropic diffusion, $\Gamma =\gamma I_n$, it follows
that $tr
F_{\bf X}$ is inversely
related to the robustness of ${\bf X}$ with respect to the onset of
diffusion noise.  In particular, for the
{\it quantum} isotropic diffusion equation
\begin{equation} \label{mastermat}
\dot{\rho} = -(\gamma/\hbar^2) \delta_{ij} [P_i, [P_j, \rho ]] - (\sigma
/\hbar^2) \delta_{ij} [X_i,
[X_j, \rho ]]
\end{equation}
(with summation over repeated indices), one finds the analog
\begin{equation}
\dot{S}_{\bf X}(0) = \gamma tr F_{\bf X} \mbox{ ,    }  
\dot{S}_{\bf P}(0) = \sigma tr F_{\bf P}
\end{equation}
of Eq. (\ref{quant}), leading to the natural generalization of Eq.
(\ref{jr})
\begin{equation} \label{jrmat}
J_r =  (\hbar/2)^{-1}[tr F_{\bf X} tr F_{\bf  P}]^{-1/2}  .
\end{equation}
for the joint robustness of a quantum state.  Comparison of Eqs.
(\ref{jointmat2}) and (\ref{jrmat}) shows
that joint nonclassicality and joint robustness are inversely related as
before.

\section{CONCLUSIONS}

It is seen that the Fisher information of a quantum observable is
essentially the variance of the nonclassical
component of the conjugate observable, as per Eqs. (\ref{vardiff}) and
(\ref{fp}).
Thus $F_X$ is a direct
measure of the nonclassicality of $P$, and vice versa.  Moreover, a
measure of joint nonclassicality for two
conjugate observables may be naturally defined to be inversely
proportional to 
the product of their Fisher
lengths, as per Eq. (\ref{joint2}).  It would be of interest to determine
whether
 pure states have a nontrivial
minimum joint nonclassicality.

Application of de Bruijn's identity to quantum diffusion processes shows
that the Fisher information of an
observable is also essentially the rate of entropy increase of the
observable at
 the onset of phase space diffusion,
and hence is inversely related to the robustness of the observable with
respect
to noise.  The Fisher length in
particular provides a {\it direct} measure of robustness, being large
when the entropy increase is small and
vice versa.  The joint robustness of two conjugate obervables is
therefore defined as being proportional to
the product of their Fisher lengths, as per Eq. (\ref{jr}).  Joint
robustness is
 simply related to joint
nonclassicality as per Eq. (\ref{inverse}):  the more nonclassical a
state is, the less robust it is, and vice
versa.

These results can be generalized to vector observables as indicated in
Sec. V, where the Fisher matrix is
essentially the nonclassical covariance matrix for the conjugate
observable, and
 its trace is essentially the rate
of entropy increase under the onset of isotropic diffusion.   Joint
nonclassicality and joint robustness are
again simply related, and indeed are inversely proportional if defined
as per Eqs. (\ref{jointmat2}) and
(\ref{jrmat}).

Finally, it is interesting to note that the
Fisher length appears naturally in a bound for the average
information which may be obtained per
measurement of $X$ on members of an ensemble ${\cal E}$.  In particular, 
this average information, $I(X|{\cal E})$, 
is given by the entropy corresponding
to the average distribution of $X$ over the ensemble
minus the average entropy of $X$ over the members of the
ensemble \cite{shan}.
From Eq. (\ref{ineq})
 one then immediately
has the bound
\begin{equation}
I(X|{\cal E}) \leq \ln \left[ (\Delta X)_{\cal E}
/(\delta X)_{\rm min}\right]
\end{equation}
i.e., the information is bounded by the logarithm of the ratio of two 
lengths (the ensemble root mean square deviation of
$X$, and  the minimum Fisher length of $X$ over the ensemble).  For higher
dimensions a corresponding
inequality may be obtained via Eq. (\ref{ineqmat}). Many other 
information-theoretic inequalities
involving Fisher information are given by Dembo et al. \cite{cov2},
and further inequalities are discussed by Romera et al. in the context of
obtaining bounds on radial expectation values \cite{romera}.

\end{document}